\documentclass[aps,longbibliography,reprint]{revtex4-1}

\usepackage[utf8]{inputenc}
\usepackage[english]{babel}
\usepackage[T1]{fontenc}
\usepackage{amsmath}
\usepackage{hyperref}

\usepackage{graphicx, float} 
\usepackage{subcaption} 
\usepackage{tabularx}
\usepackage{amssymb}
\usepackage{amsthm,dsfont}
\usepackage{natbib}

\addto\captionsenglish{}

\theoremstyle{plain}
\newtheorem{thm}{Theorem}

\newtheorem{defin}{Definition}

\newtheorem{cor}{Corollary}

\newcommand{\mc}[1]{\mathcal{#1}}

\newcommand{\bq}{\begin{equation}}
\newcommand{\eq}{\end{equation}}
\newcommand{\ba}{\begin{align}}

\newcommand\ind{\protect\mathpalette{\protect\independenT}{\perp}}
\def\independenT#1#2{\mathrel{\rlap{$#1#2$}\mkern2mu{#1#2}}}

\usepackage{color}
\definecolor{red}{rgb}{0.9,0,0}
\definecolor{green}{rgb}{0,0.8,0}
\definecolor{blue}{rgb}{0,0,0.8}
\definecolor{cautionred}{rgb}{1.0,0,0}

\definecolor{maroon}{rgb}{0.7,0,0}

\definecolor{ngreen}{rgb}{0.3,0.7,0.3}

\definecolor{golden}{rgb}{0.8,0.6,0.1}

\begin{document}

\title{Classical causal models for Bell and Kochen-Specker inequality violations require fine-tuning}
\date{\today}
\author{Eric G. Cavalcanti}
\affiliation{Centre for Quantum Computation and Communication Technology (Australian Research Council)\\
Centre for Quantum Dynamics, Griffith University, Gold Coast, Queensland 4222, Australia}
\email{e.cavalcanti@griffith.edu.au}

\begin{abstract}
Nonlocality and contextuality are at the root of conceptual puzzles in quantum mechanics, and are key resources for quantum advantage in information-processing tasks. Bell nonlocality is best understood as the incompatibility between quantum correlations and the classical theory of causality, applied to relativistic causal structure. Contextuality, on the other hand, is on a more controversial foundation. In this work, I provide a common conceptual ground between nonlocality and contextuality as violations of classical causality. First, I show that Bell inequalities can be derived solely from the assumptions of no-signalling and no-fine-tuning of the causal model. This removes two extra assumptions from a recent result from Wood and Spekkens, and remarkably, does not require any assumption related to independence of measurement settings -- unlike all other derivations of Bell inequalities. I then introduce a formalism to represent contextuality scenarios within causal models and show that all classical causal models for violations of a Kochen-Specker inequality require fine-tuning. Thus the quantum violation of classical causality goes beyond the case of space-like separated systems, and manifests already in scenarios involving single systems.

\end{abstract}

\maketitle

\section {Introduction}

Quantum contextuality, the phenomenon uncovered by Kochen and Specker (KS) \cite{Kochen1967}, is at the core of the quantum departure from classicality, and has recently been identified as a candidate for the resource behind the power of quantum computation~\cite{Howard2014}. Much controversy still exists, however, on what exactly contextuality is, with different formalisms giving different definitions of the phenomenon~\cite{Spekkens2005,Abramsky2011,Cabello2014,Acin2015}. For example, derivations following the work of Kochen-Specker require an assumption of outcome determinism, the validity of which in experimentally relevant situations has been criticised~\cite{Spekkens2005,Spekkens2014}. Indeed, it has been argued that it is not possible to experimentally test contextuality without extra assumptions~\cite{Barrett2004}.

\begin{figure}[b!]
\centering
\includegraphics[scale=1]{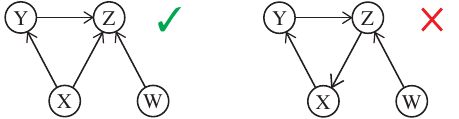}
\caption{\small In a directed acyclic graph (DAG) (left), nodes represent random variables and directed edges represent causal links. Closed cycles (right) are excluded.}
\label{DAG}
\end{figure}

Bell nonlocality~\cite{Bell1964} rests on comparatively solid foundations. It is best understood as the incompatibility between quantum correlations and causal constraints~\cite{Wiseman2017}. A modern approach is to capture these constraints within the framework of \emph{causal networks} \cite{Pearl2000}, where causal structure is represented as a \emph{directed acyclic graph} (DAG) (Fig.~\ref{DAG}). Assuming a causal graph motivated by relativity (Fig.~\ref{fig:Bell}), we can derive constraints on observable probability distributions -- the Bell inequalities. A violation of a Bell inequality thus implies either a violation of relativistic causality, or of one or more of the assumptions underlying this framework for causality, such as Reichenbach's principle of common cause~\cite{Reichenbach1956, Cavalcanti2014}. This second alternative has motivated a programme to extend the classical causal formalism to a framework of \emph{quantum causal models} \cite{Cavalcanti2014, Henson2014, Pienaar2015, Chaves2015, Costa2016, Allen2016}, opening the exciting prospect of a coherent understanding of the nature of causality in a quantum world, and a resolution of at least part of the puzzle of Bell's theorem~\cite{Cavalcanti2016}. 

Contextuality, on the other hand, is {\it a priori} unrelated to causality: it is not necessary that measurements are space-like separated, or that they involve separate subsystems at all. Thus it is not clear how a theory of quantum causality could help with contextuality. Here I bridge that gap and show that all classical causal models that reproduce the violation of Bell and KS inequalities violate a core principle of the causal models framework: {\it no-fine-tuning}. This unifies Bell-nonlocality and KS-contextuality as violations of classical causality, and opens a new direction for the study of contextuality as a resource.

In an influential work~\cite{Wood2015}, Wood and Spekkens showed that every classical causal model that can reproduce Bell inequality violations requires causal connections not observed in the phenomena, such as faster-than-light causation. Only with special, finely tuned, parameters can a causal model ``hide'' those connections from observers. This provides a novel way of looking at Bell inequalities -- not as implications of relativistic causal structure, but as implications of classical causal principles for \emph{any} causal structure. That result, however, applied only to phenomena satisfying two extra assumptions, which makes it inapplicable to contextuality scenarios. Here I prove a more general result without those extra conditions, and show how it can be extended to the case of contextuality. 

This work is organised as follows. In Section~\ref{sec:causal_models} I review the framework of causal models, and how it can be used to derive Bell inequalities by assuming \emph{free choice} and \emph{relativistic causality}. In Section \ref{sec:Meas_scenarios} I introduce a formalism for describing contextuality scenarios within classical causal models, and present the main results. In Section~\ref{sec:examples} I illustrate the need for fine-tuning in causal models for violations of Kochen-Specker inequalities with an example. In Section~\ref{sec:previous} I discuss the relation between the present result and the previous work of \cite{Wood2015}, and how it provides a novel interpretation for quantum advantage in {\it biased nonlocal games}~\cite{Lawson2010}. I conclude in Section~\ref{sec:conclusion} with a summary of the results, and a discussion of its relevance, drawbacks and directions for further research.

\section {Background}\label{sec:causal_models}

\subsection{Causal models} 

A modern framework for causation and its role in explaining correlations can be found in the theory of \emph{causal networks} \cite{Pearl2000}. With extensive applicability from statistics to epidemiology, economics, and artificial intelligence, it has been developed as a tool to connect causal inferences and probabilistic observations. In such a model, causal structure is represented as a graph $\mc{G}$, with variables as nodes and direct causal links as directed edges (arrows) between nodes. To avoid the potential for paradoxical causal loops, closed cycles are forbidden, and the resulting structure is a directed acyclic graph (DAG) (Fig.~\ref{DAG}). The relations between nodes in a DAG $\mc{G}$ can be expressed in an intuitive genealogical terminology: nodes pointing to a given node $X$ (the direct causes of $X$) are called the \textit{parents} of $X$, denoted as $Pa(X)$; the \textit{ancestors} of $X$, $An(X)$, are all nodes from which there is a directed path to $X$ (i.e. all variables in the causal past of $X$); the \textit{descendants} of $X$, $De(X)$, are all nodes for which $X$ is an ancestor (i.e. all variables in the causal future of $X$); the set of non-descendants of $X$ is denoted by $Nd(X)$.

The purpose of a DAG is to encode the conditional independences associated with any probability distribution compatible with the causal structure, through the \emph{Causal Markov Condition}: in any probability distribution $P$ that is \emph{compatible} with a graph $\mc{G}$, a variable $X$ is independent of all its non-descendants, conditional on its parents. That is, $P(X|Nd(X),Pa(X))=P(X|Pa(X)$, which we denote as $(X \ind Nd(X)|Pa(X))$. The Causal Markov Condition is equivalent to the requirement that any distribution over the variables $X_1,...,X_n$ compatible with the graph $\mc{G}$ factorises as
\begin{equation}\label{GrepresentsP}
P(X_1,...,X_n) = \prod_j P(X_j|Pa(X_j))\,.
\end{equation}

Those conditional independences can be obtained from the graph through a rule called \textit{d-separation}~\cite{Pearl2000}. Two sets of variables $X$ and $Y$ are d-separated given a set of variables $Z$ (denoted $(X\ind Y|Z)_d$) if and only if $Z$ ``blocks'' all paths $p$ from $X$ to $Y$. A path $p$ is blocked by $Z$ if and only if (i) it contains a chain $A \rightarrow B \rightarrow C$ or a fork $A \leftarrow B \rightarrow C$ such that the middle node $B$ is in $Z$, or (ii) it contains an inverted fork (head-to-head) $A \rightarrow B \leftarrow C$ such that the node $B$ is not in $Z$, and there is no directed path from $B$ to any member of $Z$.

D-separation is a \emph{sound} and \emph{complete} criterion for conditional independence: if in a DAG $\mc{G}$ two variables $X$ and $Y$ are d-separated given $Z$, $(X \ind Y|Z)_d$, then they are conditionally independent given $Z$, $(X \ind Y|Z)$, in all distributions compatible with $\mc{G}$; and if for all distributions compatible with $\mc{G}$, the conditional independence $(X \ind Y|Z)$ holds, then $\mc{G}$ satisfies $(X \ind Y|Z)_d$.

\subsection{Causal models and Bell's theorem} \label{sec:Bell_models}

\begin{figure}[h]
\includegraphics[scale=1]{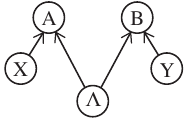}\centering
\caption{\small DAG representing the causal structure of a Bell scenario.}
\label{fig:Bell}
\end{figure}

As an example of the application of this framework, we review how it can be used to derive Bell's theorem.  Consider the correlations between measurements performed by two agents, Alice and Bob. Alice's choice of measurement is represented by a variable $X$, and Bob's by a variable $Y$. Their respective outcomes are represented by $A$ and $B$. Their measurements are assumed to be performed within space-like separated regions, so that no relativistic causal connection can exist between the variables in Alice's lab and those in Bob's lab. They may however be correlated due to variables in their common causal past, the set of which is denoted by $\Lambda$. The assumption that Alice and Bob can make ``free choices'' is translated as the requirement that $X$ and $Y$ are \emph{exogenous} variables: they have no relevant causes. This scenario is represented in the graphical notation as in Fig.~\ref{fig:Bell}.

The Causal Markov Condition then implies that $P(AB\Lambda|XY)=P(A|X\Lambda)P(B|Y\Lambda)P(\Lambda)$. Averaging over $\Lambda$ we obtain the factorisability condition of a local hidden variable model:
\begin{equation}
P(AB|XY) = \sum_\Lambda P(\Lambda)P(A|X\Lambda)P(B|Y\Lambda) \, .
\end{equation}

As is well known, this leads to the Bell inequalities, which can be violated by quantum correlations~\cite{Brunner2014}. Therefore, assuming relativistic causal structure and free choices, quantum correlations cannot be reproduced by the classical framework of causality.

Note that the assumption of ``free choice'' is not strictly necessary: a weaker but sufficient condition is simply ``$\Lambda$-independence'', that the measurement choices are independent of any latent variables $\Lambda$ that are causally connected with the systems, $(\Lambda\ind XY)$. This would still be compatible with a causal graph where there is a common cause between the measurement choices $X$ and $Y$. But without one of these assumptions, it would be possible for a conspiratorial ``superdeterministic'' theory to reproduce the quantum correlations. Remarkably, in the main result of this paper {\it neither} of these assumptions is needed. Superdeterministic theories, for example, are ruled out because they violate no-fine-tuning.

\section{Contextuality scenarios and causal models} \label{sec:Meas_scenarios}

\subsection{Basic definitions} 

Traditionally, locality and noncontextuality have been spelled out in terms of ontological models, but here I translate those concepts into the language of causal models. Indeed, we can think of ontological models as causal models in disguise, which is an useful perspective as it allows one to more clearly identify implicit classical causal assumptions that may be revised in light of quantum causal models. The formalism used here is most analogous to that of Abramsky and Brandenburger \cite{Abramsky2011}, although it is expressed in the language of causal models, and uses simplified terminology (for example, without reference to sheaf theory).

A \emph{measurement scenario} is specified by: i) A set $\mc{M}$ of measurements; ii) for each measurement $m \in \mc{M}$, a set of possible outcomes $\mc{O}_m$; and iii) a \emph{compatibility structure} $\mc{C}$ on $\mc{M}$ -- a family of subsets of $\mc{M}$ -- specifying joint measurability. Two measurements $m_1,m_2\in \mc{M}$ are said to be \emph{jointly measurable}, \emph{compatible}, or to be part of a \emph{context}, iff $\{m_1,m_2\}\in\mc{C}$, and likewise for sets of more than two measurements~\footnote{Note that pairwise compatibility in a set of measurements $\mc{M}$ does not in general imply that all measurements in $\mc{M}$ are jointly measurable~\cite{Liang2011}.}. Without loss of generality, we can enlarge each $\mc{O}_m$ so that all measurements have the same number of outcomes, and label them so that all outcome sets are equal and denoted simply by $\mc{O}$. 

Consider now an individual test within a measurement scenario, where a set of $n$ random variables $X_1,...,X_n$ specifies $n$ measurements to be performed upon the system. A \emph{compatibility scenario}, or \emph{contextuality scenario} is one in which for any given run $x_1,...,x_n$ are jointly measurable. That is, let $X_i=x_i\in\mc{M}$ denote the values of those variables in a particular run. Then $\{x_1,...,x_n\}\in\mc{C}$. This could be done, for example, through a further random variable $C$ that selects a context, from which $X_1,...,X_n$ are determined, but we make no assumption about the process of selection or the order in which the measurements are performed. A special class of contextuality scenarios is that in which $\mc{M}$ can be decomposed into $k$ subsets $\mc{M}_1, \mc{M}_2,...,\mc{M}_k$ such that all contexts $c\in\mc{C}$ have at most one element from each subset. These are called ($k$-partite) \emph{Bell-nonlocality scenarios}. 

From here on we consider scenarios containing only \emph{pairs} of compatible measurements. This doesn't necessarily mean that there are no sets of three or more compatible measurements that can be performed on the system, but only that we are restricting $\mc{M}$ so that it only contains pairs of them. These will be called \emph{binary} contextuality scenarios, a special case of which are bipartite Bell scenarios. The measurements will be chosen through random variables $X$ and $Y$, with outcomes respectively recorded by random variables $A$ and $B$. 

A \emph{phenomenon} $\mc{P}$ for such a scenario is specified by a probability distribution $\mc{P}(ABXY)$ for the observable variables. Note that no causal assumption is made up to this stage. We now define what we mean by a (classical) causal model for a phenomenon. 

\begin{defin}[Causal model] \label{defin:CM}
A (classical) \emph{causal model} $\Gamma$ for a phenomenon $\mc{P}$ consists of a (possibly empty) set of latent variables $\Lambda$, a DAG $\mc{G}$ with nodes $\{A,B,X,Y,\Lambda\}$, and a probability distribution $P(ABXY\Lambda)$ compatible with $\mc{G}$, such that $\mc{P}(ABXY)=\sum_\Lambda P(ABXY\Lambda)$.
\end{defin}

A special class of phenomena are those where the probability for the outcome of one measurement does not depend on which measurement it may be performed together with, i.e.~that satisfy the property of \emph{no-disturbance}.
\begin{defin}[No-disturbance] \label{defin:ND}
A phenomenon is said to satisfy \emph{no-disturbance} iff $\mc{P}(A|XY)=\mc{P}(A|X)$ and $\mc{P}(B|XY)=\mc{P}(B|Y)$ for all values of the variables $A$, $B$, $X$, $Y$ for which those conditionals are defined.
\end{defin}
In the causal-model notation, the no-disturbance conditions are denoted by $(A\ind Y|X)$ and $(B\ind X|Y)$. Compatibility scenarios naturally satisfy those conditions, since $X$ and $Y$ can only take joint values as pairs of compatible measurements, and the no-disturbance condition is implicit in the very meaning of compatibility~\cite{Abramsky2011}. In Bell scenarios this assumption is called \emph{no-signalling}, and where the measurements $X$ and $Y$ are performed in space-like separated regions, it is justified by relativity.

Whereas no-disturbance and no-signalling are purely properties \emph{of phenomena}, the subsequent definitions are about properties \emph{of causal models} for phenomena in contextuality scenarios. We say that a phenomenon violates a property when no causal model for the phenomenon satisfies that property.

Bell-locality and KS-noncontextuality are equivalent to the existence of a factorisable hidden variable model, which in the language of causal models translates to: 

\begin{defin}[Factorisability] \label{defin:FACT}
A causal model is said to satisfy \emph{factorisability} iff $\forall \,A,B,X,Y,$ $P(AB|XY)=\sum_\Lambda P(\Lambda) P(A|X\Lambda) P(B|Y\Lambda)$.
\end{defin}
The original Kochen-Specker theorem assumed the existence of a deterministic noncontextual model. By Fine's theorem~\cite{Fine1982}, this is equivalent to the existence of a factorisable model, and to the existence of a joint probability distribution for the outcomes of all measurements that returns the observable correlations as marginals.
\begin{defin}[KS-noncontextuality] \label{defin:KS-NC}
A causal model for a contextuality scenario is said to satisfy \emph{KS-noncontextuality} iff it is factorisable.
\end{defin}
From factorisability, one can derive, for each contextuality scenario, inequalities that bound the set of KS-noncontextual phenomena, as facets of a convex polytope \cite{Abramsky2012, Pitowsky1989}. These are the \emph{KS-inequalities}~\footnote{A note on terminology: some authors (e.g.~\cite{Bengtsson2012}) use ``NC inequalities''~\cite{Klyachko2008,Cabello2009} to refer to what I call KS inequalities, and use the latter term to refer to inequalities which, apart from factorisability, make the extra assumption that outcomes assigned by the hidden variables satisfy the same constraints as the corresponding quantum operators (e.g., that outcomes corresponding to orthogonal projectors are mutually exclusive)~\cite{Simon2001,Larsson2002}. However, for every ``KS inequality'' (in the sense of those authors) that has a quantum violation, there exists a ``NC inequality'' that is violated by the same statistics. The situation is analogous to that of the original Bell inequality~\cite{Bell1964}, which apart from a local hidden variable model, also assumed a quantum prediction of perfect correlations. The term ``Bell inequality'', however, now refers to those derived without that unnecessary assumption. Likewise, I advocate the use of ``KS inequalities'' to refer to those derived without assuming the quantum predictions.}, which reduce to \emph{Bell inequalities} in Bell scenarios.

Despite this formal equivalence. much controversy remains regarding the justification for factorisability in contextuality scenarios. While in Bell scenarios it is implied by Bell's notion of local causality~(for a detailed review, see \cite{Wiseman2017}), this is not so in contextuality scenarios, where measurements are not space-like separated. To derive factorizability in this case, one requires, apart from the assumption of measurement noncontextuality, an extra assumption of outcome determinism~\cite{Spekkens2014}, calling into question the implication of violations of Kochen-Specker inequalities. However, as we will see below, factorisability is implied by another fundamental principle of the causal models framework, the principle of \emph{no-fine-tuning}, or \emph{faithfulness}. 
\begin{defin}[Faithfulness (no fine-tuning)] \label{defin:faithfulness}
A causal model $\Gamma$ is said to satisfy \emph{no fine-tuning} or be \emph{faithful} relative to a phenomenon $\mc{P}$ iff every conditional independence $(C\ind D|E)$ in $\mc{P}$ corresponds to a d-separation $(C\ind D|E)_d$ in the causal graph $\mc{G}$ of $\Gamma$.
\end{defin}
For example, suppose a phenomenon satisfies the no-disturbance condition $(A\ind Y|X)$, but its causal structure contains a direct link from $Y$ to $A$, and thus lacks the d-separation $(A\ind Y|X)_d$. This is only possible if some of the parameters of the model take special values that hide the influence of $Y$ on $A$. An example is Bohmian mechanics, where the probability distributions for the hidden variables are constrained by the ``quantum equilibrium'' condition \cite{Durr1992}. A faithful causal model, on the other hand, has no such hidden causal connections.

No-fine-tuning can be seen therefore as an instance of Occam's razor~\cite{Pearl2000}, or of a methodological principle that also motivates Spekkens' notion of generalised contextuality~\cite{Spekkens2005}: {\it Leibniz's principle of the identity of indiscernibles}. It states that one should not postulate differences in the ontological (i.e., causal-model) description of a phenomenon where none exists at the operational level. Here, however, we show that this principle leads to KS-noncontextuality, implying that the two competing notions of noncontextuality have more in common than has been previously realised. Whether Spekkens' notion can also be directly derived from faithfulness is an interesting question left for future work.

\subsection{Main result}\label{sec:main_result}

\begin{thm} \label{thm:main} Every faithful causal model for a no-disturbance phenomenon is factorisable.
\end{thm}

As shown in detail in the Appendix, the proof of Theorem~\ref{thm:main} proceeds by showing that all DAGs that do not require fine-tuning to explain the no-disturbance conditions lead to factorisability. We first note that the no-disturbance conditions, together with the assumption of no fine-tuning, imply that every DAG $\mc{G}$ for a no-disturbance phenomenon $\mc{P}$ must satisfy the d-separation conditions $(A\ind Y|X)_d$ and $(B\ind X|Y)_d$. We thus proceed by excluding every DAG that does not satisfy these conditions, and show that all remaining DAGs imply factorisability. As an immediate corollary:
\begin{cor} \label{cor:KSNC} No fine-tuning and no-disturbance imply KS noncontextuality.
\end{cor}

Another corollary is a stronger version of the result of \cite{Wood2015}, without the extra assumptions of marginal setting independence and local setting dependence:
\begin{cor} No fine-tuning and no-signalling imply Bell-locality.
\end{cor}

It is instructive to state Corollary \ref{cor:KSNC} in a contrapositive form:
\begin{cor} Every classical causal model that reproduces the violation of a KS-inequality in a no-disturbance phenomenon requires fine-tuning.
\end{cor}

This result implies that the quantum violation of classical causality, long recognised in the case of space-like separated entangled quantum systems, also manifests in the case of single systems.

\section{Examples}\label{sec:examples}

To illustrate the need for fine-tuning in causal explanations of KS-inequality violations, let us consider as a simple example the three-observable scenario introduced Liang, Wiseman and Spekkens (LSW) in \cite{Liang2011}. The measurement scenario consists of a set of three measurements $\mathcal{M}=\{m_1,m_2,m_3\}$ with binary outcomes $\mathcal{O}=\{0,1\}$ and compatibility structure $\mathcal{C}=\{ \{m_1,m_2\}, \{m_1,m_3\}, \{m_2,m_3\}\}$ -- that is, they are pairwise compatible but not triplewise compatible. In any given experimental test, $X$ and $Y$ can take any pair of values from $\mathcal{C}$ and $A$ and $B$ can take values from $\mathcal{O}=\{0,1\}$. Consider a phenomenon $\mathcal{P}(ABXY)$ that satisfies the no-disturbance relations $\mc{P}(A|XY)=\mc{P}(A|X)$ and $\mc{P}(B|XY)=\mc{P}(B|Y)$. Thus from Theorem~\ref{thm:main} any faithful causal model for this phenomenon must satisfy KS-noncontextuality.

KS-noncontextuality in this scenario implies the LSW inequality~\cite{Liang2011}:
\begin{equation}\label{eq:LSW}
\sum_{\{m_i,m_j\}\in\mathcal{C}}\frac{1}{3}\mathcal{P}(A\neq B|m_i,m_j) \leq \frac{2}{3} \,.
\end{equation}
To see this, recall that the existence of a KS-noncontextual model is equivalent to the existence of a joint probability distribution for the outcomes of all measurements. For each extreme point of this distribution, at most two of the three pairs of measurements can be anti-correlated, so the average probability of anti-correlation cannot exceed 2/3. (Note that since $(X,Y)=(m_1,m_2)$ and $(X,Y)=(m_2,m_1)$ correspond to the same pair of measurements being performed, $\mathcal{P}(A=k,B=l|X=m_i,Y=m_j)=\mathcal{P}(A=l,B=k|X=m_j,Y=m_i)$). 

The authors of \cite{Liang2011} illustrate a hypothetical maximal violation of \eqref{eq:LSW} as a realisation of the ``parable of the overprotective seer'' (OS): wishing to ward off unworthy suitors for his beloved daughter, a seer from ancient Assyria proposed to each of them the following task. They were taken to a table upon which sat three boxes, and asked to open two of the boxes. If both boxes were to contain a gem, or both boxes were to not contain a gem, they would win the task and would be allowed to marry the seer's daughter. As it turned out, every suitor always randomly found a gem in one box and none in the other -- a seemingly paradoxical situation!

\begin{figure}[h]
\centering
\includegraphics[scale=1]{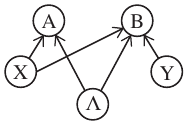}
\caption{\small Example of a causal structure that can reproduce violations of KS-inequalities with fine-tuning.}
\label{fig:FT_DAG}
\end{figure}

An example of a causal model that realises the OS correlations consists of the causal graph in Figure~\ref{fig:FT_DAG} and the following causal parameters. A uniformly distributed latent variable $\Lambda$ determines for each box whether or not it contains a gem. Let $X$ denote the first box opened by the suitor, so $A(X,\Lambda) \in \{ 0,1\}$ for any value of $X$, with $A=1$ representing the presence of a gem. Depending on whether or not the first box contained a gem, a gem is added or removed from the second box $Y$, to ensure that $B(X,Y,\Lambda)=A(X,\Lambda)\oplus 1$ for all $\{X,Y\}\in\mathcal{C}$. To see the need for fine-tuning, note that if one were able to prepare any distribution for the hidden variables, no-disturbance would be violated. For example, suppose one could prepare $\Lambda=\Lambda_0$ such that $A(X=m_i,\Lambda_0) = \delta_{i,1}$. If $Y=m_3$ is measured, its outcome will be $B=0$ if $X=m_1$ and $B=1$ if $X=m_2$. Of course, the need for fine-tuning would manifest in different ways for other causal models but Theorem~\ref{thm:main} guarantees that any causal model for these correlations requires fine-tuning.

Although the LSW inequality does not have a quantum violation for projective measurements, a very similar analysis would hold for the slightly more complex scenario of Klyachko {\it et al.} \cite{Klyachko2008}, involving five observables with binary outcomes, and cyclical compatibility structure. We leave this as an exercise for the reader.

\section{Relation to previous work}\label{sec:previous}

The present work was inspired by \cite{Wood2015}, however, there are some important differences. In \cite{Wood2015}, it was shown that any causal model for a Bell scenario that satisfies the assumptions of no-signalling and no-fine-tuning, plus two extra assumptions of \emph{marginal setting independence} and \emph{local setting dependence}, is factorisable. As the present derivation does not make those extra assumptions, it generalises the result of \cite{Wood2015} already for Bell scenarios.

Local setting dependence is the assumption that the local outcome is \emph{not} independent of the local setting, i.e., that it is not the case that $(A\ind X)$ or $(B\ind Y)$. This assumption rules out standard examples of Bell-inequality violation with unbiased outcomes. Marginal setting independence is the assumption that the settings $X$ and $Y$ are uncorrelated, $(X\ind Y)$. While in \cite{Wood2015} the no-signalling condition is justified by space-like separation, here the more general `no-disturbance' is implied by measurement compatibility. In Bell scenarios, the compatibility of $X$ and $Y$ is guaranteed because they are chosen from disjoint sets of measurements $\mc{M}_A$ and $\mc{M}_B$, but in general contextuality scenarios $X$ and $Y$ are chosen from the same set $\mc{M}$, and are not in general independent. Thus the result of \cite{Wood2015} cannot be applied to contextuality scenarios. 

Note that the Bell inequalities implied by the present result are the usual constraints on conditional probabilities $\mc{P}(AB|XY)$. In some contexts, such as that of biased nonlocal games~\cite{Lawson2010}, one may be interested in constraints on the unconditional $\mc{P}(ABXY)$. The class of classical models considered in that paper is one where the usual constraints on conditional probabilities apply, but where the joint probabilities for $X$ and $Y$ are allowed to be arbitrary. Therefore $\mc{P}(ABXY)=\mc{P}(XY)\mc{P}(AB|XY)=P(XY)\sum_\Lambda P(\Lambda) P(A|X\Lambda) P(B|Y\Lambda)$. It was shown in \cite{Lawson2010} that quantum correlations can violate inequalities implied by this model for some values of $\mc{P}(XY)$, even in cases where the conditional probabilities do not violate a Bell inequality. Since here we make no assumption about the joint probabilities $\mc{P}(XY)$, the set of classical models above can also be derived from the assumption of no-fine-tuning, and therefore classical causal models that reproduce the quantum violations of those constraints also require fine-tuning -- even when they do not allow for the violation of a Bell inequality.

\section{Conclusion}\label{sec:conclusion}

In summary, we have derived KS-inequalities as a consequence of the principle of no-fine-tuning for causal models of phenomena that satisfy the no-disturbance conditions. This result unifies Bell nonlocality (where `no-disturbance' corresponds to `no-signalling') and KS-contextuality as violations of classical causality. Remarkably, unlike all other derivations of Bell inequalities, this result needs no assumption related to independence of settings, such as `free choice', `$\Lambda$-independence', or `marginal setting independence'. 

One could object that to achieve the no-disturbance conditions in general contextuality scenarios, one requires perfectly compatible measurements, and that this idealisation makes it inapplicable to real experimental tests. This is true, but it is no worse than the problem faced by all standard derivations of KS-inequalities, as discussed in \cite{Spekkens2014}, where the assumption of outcome determinism is incompatible with realistic unsharp measurements. The advantage of the present work is that, at least for idealised phenomena, it allows the derivation of KS-inequalities from causality principles alone, without the extra assumption of outcome determinism. Thus it allows for the conclusion that those causal principles must be revised in light of the (idealised) predictions of quantum theory, whereas no such conclusion can be reached with the usual derivation, even in the idealised case.

Furthermore, the present derivation suggests a path for an experimentally robust generalisation. Given a \emph{measure} of causal connection \cite{Janzing2013, Chaves2015a}, we can propose a \emph{generalised principle of no-fine-tuning}: a causal model should not allow causal connections \emph{stronger than needed} to explain the observed deviations from no-disturbance. It would be interesting to determine whether testable constraints can be derived this way. Another interesting question is whether the proof can be extended to arbitrary numbers of measurements per context.

From the point of view of applications, we may understand fine-tuning as a kind of ``resource waste'', postulating causal links that are not directly observed in the phenomena, but are washed-out by our ignorance of underlying parameters. It is plausible to conjecture that quantum causal models \cite{Henson2014, Pienaar2015, Chaves2015, Costa2016, Allen2016}, on the other hand, can avoid fine-tuning in explaining quantum  correlations. If we think of classical causal models as classical simulations of quantum phenomena, the present result implies that they must necessarily waste resources via fine-tuning. A quantification of this intuition could potentially provide a novel picture to explain the power of contextuality as a resource for quantum computation. 

Finally, it would be interesting to determine whether the generalised notions of noncontextuality given by the formalism of Spekkens \cite{Spekkens2005} can also be understood as arising from no-fine-tuning.

\begin{acknowledgments} This research was supported by Grant No. FQXi-RFP-610 1504 from the Foundational Questions Institute Fund (fqxi.org) at the Silicon Valley Community Foundation, and by the Australian Research Council Centre of Excellence project number CE170100012. The author would like to acknowledge helpful discussions with Robert Spekkens, Nadish de Silva, Howard Wiseman, Rafael Chaves, Leandro Aolita, Sally Shrapnel, Fabio Costa, David Schmid and Ad\'{a}n Cabello, and constructive comments from two anonymous referees.
\end{acknowledgments}

%

\appendix

\section{Proof of Theorem 1}

\begin{proof}
We prove by exhaustion that all DAGs that do not require fine-tuning to explain the no-disturbance conditions lead to factorisability, and thereby to KS-noncontextuality.

First note that the no-disturbance conditions, together with the assumption of no fine-tuning, imply that every DAG $\mc{G}$ for a no-disturbance phenomenon $\mc{P}$ must satisfy the d-separation conditions $(A\ind Y|X)_d$ and $(B\ind X|Y)_d$. We thus proceed by excluding every DAG that does not satisfy these conditions, and showing that all remaining DAGs imply factorisability.

The class of DAGs we need to consider are those that include latent variables as common causes for observable variables, or direct causal connections between variables. There is no point considering latent variables as intermediaries between variables, or as common effects of variables, since adding those has no effect on the allowed probability distributions over the observable variables.

To aid the proofs, we introduce the graphical notation in Figs.~\ref{fig:DAGnotation-1}-\ref{fig:DAGnotation2} to represent sets of causal connections.
\begin{figure}[!h]
	\includegraphics[scale=1]{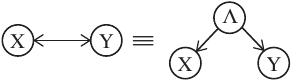}\centering\\
	\includegraphics[scale=1]{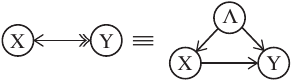}\centering
	\caption{\small Shortcut graphical notations for causal connections between $X$ and $Y$.}
	\label{fig:DAGnotation-1}
\end{figure}
\begin{figure}[!h]
	\includegraphics[scale=1]{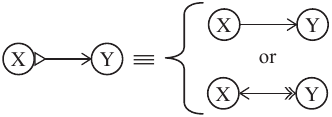}\centering
	\caption{\small Shortcut graphical notation for a direct cause from $X$ to $Y$ with or without a common cause.}
	\label{fig:DAGnotation0}
\end{figure}
\begin{figure}[!h]
	\includegraphics[scale=1]{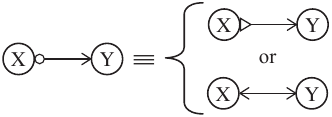}\centering
	\caption{\small Shortcut graphical notation for any causal link with no direct cause from $Y$ to $X$.}
	\label{fig:DAGnotation1}
\end{figure}
\begin{figure}[!h]
	\includegraphics[scale=1]{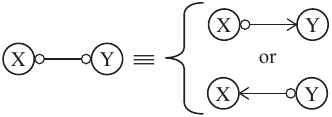}\centering
	\caption{\small Shortcut graphical notation for any causal link between $X$ and $Y$.}
	\label{fig:DAGnotation2}
\end{figure}

As a reminder, the d-separation condition $(A\ind Y|X)_d$ means that $X$ ``blocks'' all paths $p$ from $A$ to $Y$. A path is blocked by $X$ if and only if (i) it contains a chain or a fork with $X$ in the middle node, or (ii) it contains an inverted fork such that $X$ is not the middle node nor a descendant of it. The intuition behind these rules is that in case (i), $X$ is a common cause or an intermediate cause between its adjacent variables, and thus conditioning on $X$ eliminates the correlations established by this causal path, thus ``blocking'' it. In case (ii), $X$ is a common effect of its adjacent variables, and thus conditioning on $X$ can render them correlated. From this d-separation condition therefore we eliminate all graphs that contain one or more paths between $A$ and $Y$ that are not blocked by $X$, and likewise for $(B\ind X|Y)_d$. 

\emph{Step 1:} From the d-separation condition $(A\ind Y|X)_d$, we can exclude any direct causal link or common cause between $A$ and $Y$ (i.e.~all edges of the kind shown in Fig.~\ref{fig:DAGnotation2}). Likewise from $(B\ind X|Y)_d$, we can exclude any direct causal link or common cause between $B$ and $X$. Taken together, these exclude common causes between any three or all four of the variables. We are left with the possibility of any causal links between the pairs $\{A,B\},$ $\{X,A\},$ $\{Y,B\}$ and $\{X,Y\}$, as shown on Fig.~\ref{fig:ProofStep1}. Note that we do not assume a priori that common causes can only act between at most two variables -- this is a consequence of no-fine-tuning. 
\begin{figure}[!h]
\includegraphics[scale=1]{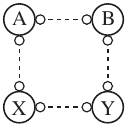}\centering
\caption{\small Remaining class of DAGs after step 1. Dashed lines represent a connection of the type indicated or its absence.}
\label{fig:ProofStep1}
\end{figure}

\emph{Step 2a:} Next, we exclude a direct causal link from $A$ to $B$ (with or without a common cause between those two variables). First, note that the assumption of such a link excludes any causal link between $A$ and $X$, as those would violate $(B\ind X|Y)_d$, and a direct link from $B$ to $Y$, as this would violate $(A\ind Y|X)_d$. The remaining class of graphs compatible with a direct link from $A$ to $B$ can now have any connection between $X$ and $Y$ plus any link with no direct cause from $B$ to $Y$ (Fig.~\ref{fig:ProofStep2a}).
\begin{figure}[!h]
\includegraphics[scale=1]{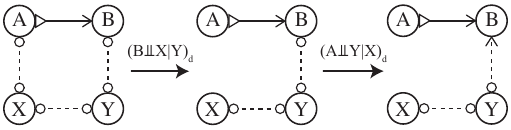}\centering
\caption{\small DAG elimination from d-separation in step 2a. The diagrams in this and the following figures should be read as follows. Each graph represents the set of all DAGs that contain only causal connections compatible with each shortcut notation. Solid lines require a compatible causal connection, and dashed lines are compatible with no connection. On each arrow between graphs, we eliminate all graphs that are incompatible with the d-separation condition indicated.}
\label{fig:ProofStep2a}
\end{figure}

\emph{Step 2b:} We now exclude a common cause between $B$ and $Y$ acting together with a direct link from $X$ to $Y$ and/or a common cause between $X$ and $Y$; those graphs would violate $(B\ind X|Y)_d$ as they are colliders. There are now two classes of graphs compatible with a direct link between $A$ and $B$: i) any link between $X$ and $Y$ and a direct link from $Y$ to $B$; or ii) a direct link from $Y$ to $X$ plus any link with no direct cause from $B$ to $Y$.(Fig.~\ref{fig:ProofStep2b}). 
\begin{figure}[!h]
\includegraphics[scale=1]{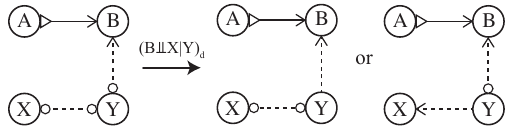}\centering
\caption{\small DAG elimination from d-separation in step 2b.}
\label{fig:ProofStep2b}
\end{figure}

\emph{Step 2c:} We proceed to show that all phenomena compatible with the two classes of graphs remaining after Step 2b are factorisable. To see this, first note that both classes of graphs i) and ii) above respect the following d-separation conditions: $(AB\ind X|Y)_d$ and $(A\ind Y)_d$. This means that all distributions compatible with those graphs must respect $P(AB|XY)=P(AB|Y)$ and $P(A|Y)=P(A)$. The first conditional independence implies that the joint distribution of $A$ and $B$ doesn't depend on the choice of measurement $X$, which intuitively should imply that this phenomenon cannot be contextual. To see this formally, note that from the definition of conditional probability and the two equations above we get $P(AB|XY)=P(B|AY)P(A|Y)=P(B|AY)P(A)$. Now let $\Lambda$ be a variable that determines $A$, so that $P(A) = \sum_\Lambda P(\Lambda) P(A|\Lambda)$ and $P(B|AY) = P(B|Y\Lambda)$. Then $P(AB|XY)=\sum_\Lambda P(\Lambda) P(A|\Lambda)P(B|Y\Lambda)$, which is a factorisable model with no dependence on $X$. 

This concludes the part of the proof excluding a direct causal link from $A$ to $B$. By symmetry we exclude any direct causal link from $B$ to $A$. The remaining class of graphs now can have a common cause between $A$ and $B$ and any causal link between the pairs $\{X,A\},$ $\{Y,B\}$ and $\{X,Y\}$ (Fig.~\ref{fig:ProofStep2c}).

\begin{figure}[!h]
\includegraphics[scale=1]{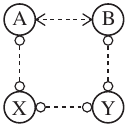}\centering
\caption{\small Remaining class of DAGs after step 2.}
\label{fig:ProofStep2c}
\end{figure}

\emph{Step 3:} We now proceed to exclude, from the remaining graphs, any direct cause from $A$ to $X$ (a retrocausal model). First we see that, assuming such link, a common cause between $A$ and $B$ is excluded from $(B\ind X|Y)_d$. Next, any link between $X$ and $Y$ except $X \rightarrow Y$ is excluded from $(A\ind Y|X)_d$. Finally, any link between $Y$ and $B$ except $Y \rightarrow B$ is excluded by $(B\ind X|Y)_d$. 

The remaining graph has four possible links: a common cause between $A$ and $X$, $A \rightarrow X$, $X \rightarrow Y$ and $Y \rightarrow B$. This implies that $(B\ind AX|Y)$ and $(A\ind Y|X)$. Thus $P(AB|XY)=P(B|AXY)P(A|XY)=P(A|X)P(B|Y)$, which is trivially factorisable. By symmetry, we eliminate any direct cause from $B$ to $Y$.
\begin{figure}[h]
\includegraphics[scale=1]{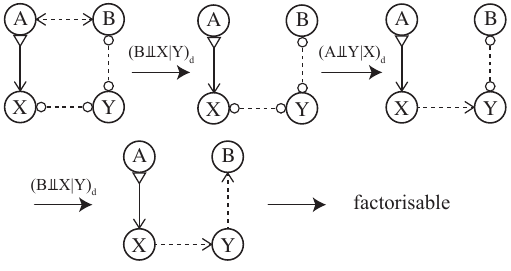}\centering
\caption{\small Elimination of DAGs in step 3.}
\label{fig:ProofStep3}
\end{figure}

\emph{Step 4:} After step 3 we are left with the following class of graphs (Fig.~\ref{fig:ProofStep3left}): a possible common cause (let's call it $\Lambda$) between $A$ and $B$, any link between $X$ and $Y$, no direct cause $A \rightarrow X$ and no direct cause $B \rightarrow Y$.
\begin{figure}[h]
\includegraphics[scale=1]{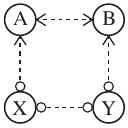}\centering
\caption{\small Remaining class of DAGs after step 3.}
\label{fig:ProofStep3left}
\end{figure}

In the next step we exclude, from $(A\ind Y|X)_d$, any common cause $A \leftrightarrow X$ acting together with $X \leftrightarrow Y$ and/or $X \leftarrow Y$. Likewise from $(B\ind X|Y)_d$, we exclude any common cause $B\leftrightarrow Y$ acting together with $X \leftrightarrow Y$ and/or $X \rightarrow Y$.

\begin{figure}[H]
\includegraphics[scale=1]{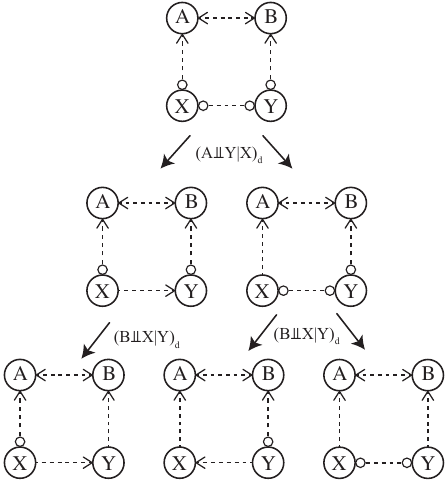}\centering
\caption{\small DAG elimination in step 4. In step 5 we show that the three remaining classes of DAGs are factorisable.}
\label{fig:ProofStep4}
\end{figure}

\emph{Step 5:} We are finally left with three remaining classes of graphs. All three classes allow for $A \leftrightarrow B$, $X \rightarrow A$ and $Y \rightarrow B$, and respectively i) $X \leftrightarrow A$, $X \rightarrow Y$; ii) $Y \leftrightarrow B$ $X \leftarrow Y$; and iii) any link between $X$ and $Y$. All of these have $\Lambda$ as a free variable and imply the conditional independences $(A\ind BY|X\Lambda)$ and $(B\ind AX|Y\Lambda)$. So $P(AB|XY)=\sum_\Lambda P(\Lambda) P(A|BXY\Lambda)P(B|XY\Lambda)=\sum_\Lambda P(\Lambda) P(A|X\Lambda)P(B|Y\Lambda)$, which is of factorisable form. This completes the proof.

\end{proof}

\end{document}